\documentstyle[preprint,aps]{revtex}
\begin{document}
\title{Soliton and 2D domains in ultra-thin magnetic films}
\author{ S. T. Chui$^a$ and V. N. Ryzhov$^b$}
\address{a: Bartol Research Institute, University of Delaware, Newark, DE
19716\\b:Institute for High Pressure Physics, Russian Academy of
Sciences, 142 092 Troitsk, Moscow region, Russia}
\maketitle
\begin{abstract}
We show that many {\bf two dimensional} domain patterns observed in
Monte Carlo simulations can be obtained from the {\bf many} soliton solutions
of the imaginary time Sine Gordon equation. This opens the door to
analytic physical understanding of the micromagnetics in ultra-thin
films.
\end{abstract}
There has been much experimental
interest recently in the magnetism of ultra-thin films.\cite{Prinz,Heinrich}
partly motivated by the possible
integration of the semi-conductor microelectronics technology
with magnetic elements\cite{Prinz} and possible device applications with
the giant magnetoresistive (GMR) effect. From a fundamental
physics viewpoint,
these systems present opportunities for studying new phenomena that
are beginning to be uncovered.
The interaction energy between the spins at positions ${\bf R}$, ${\bf
R}'$ is
\begin{equation}
H=0.5 \sum_{ij=xyz,{\bf R}{\bf R}'} V_{ij}({\bf R}-{\bf R}')S_i({\bf R})S_j({\bf R}')
\end{equation}
where $V=V_d+V_e+V_a$ is the sum of the  dipolar energy
$V_{dij}({\bf R})=g\nabla_i\nabla_j (1/|{\bf R}|);$
the exchange energy $V_e=-J\delta({\bf R}={\bf R}'+d)\delta_{ij};$ and the
crystalline anisotropy energy $V_a.$ Here $d$ denotes
the nearest neighbours. $g$ and $J$ are coupling constants.
The form of the anisotropy energy depends on the material of interest. It
can be uniaxial (e.g. $V_a=-K\sum_iS_{ix}^2$)
or four-fold symmetric (e.g. $V_a=-K\sum_i[S_{ix}^2-S_{iy}^2]^2/4$),
with the easy or hard axis aligned along specific directions.
The dipolar interaction often lead to the formation of domains.
The pattern of the domains have recently received
considerable interest under the context of the ``self-assembled''
systems where ${\bf electric}$ dipoles lead to the formation of
domains in Langmuir films. Whereas the electric
dipoles are always perpendicular
to the film plane in that case, the magnetic dipoles can be parallel or
perpendicular to the plane\cite{Heinrich,magz,pb,ber1,ber2}.
For discussions in this
paper, we restrict our attention to those cases so that the spins lie
in the plane of the film, the case of experimental interest
in sensor type applications.

The domain pattern depends on the shape of the sample,
which is especially important for small structures. The
physics of the pattern of domains in small magnetic structures
is the subject of the present paper.
We have been studying the physics
of spin reversals of different small structures\cite{mikonos},
such as monolayer films with perpendicular\cite{nucl} and 4-fold in
plane\cite{nuclang} anisotropy,
nanowires and particles\cite{part}, coupled films\cite{2l}
and the shape of the nucleus\cite{ellipse}. This paper reports our
findings that much of the
domain patterns observed in the numerical simulations
can be reproduced as the {\bf analytic many soliton} solutions of
the imaginary time Sine-Gordon equation. This is illustrated by two
examples in Fig 1 and 2 where we show the simulation and
analytic results side by side.
These analytic results have the potential for greatly improving understanding
quantitatively the domain structure and the switching process in small
structures. Thus analytic calculations can be performed to predict
trends as the system parameters are changed. These analytic results can be
used as a starting point of a simulation, considerably shortening the
simulation time; sometimes the simulations become entirely unnecessary.
We now explain our results in detail.

Mathematically in the continuum approximation, the dipolar energy
$E_d\approx \frac{g}{2}\int d{\bf R}d{\bf R}'S_i({\bf R})S_j({\bf
R}')\nabla_i\nabla_j(1/|{\bf R-R'}|)$
can be written in terms of the
magnetic charges $\nabla\cdot {\bf S}$
after two integration by parts and neglecting the
boundary terms as
$E_d\approx \frac{g}{2}\int d{\bf R}d{\bf R}'\nabla\cdot {\bf S}({\bf R})\nabla
\cdot {\bf S}({\bf R}')(1/|{\bf R-R'}|).$
Thus the dipolar energy is reduced if the ``magnetic charges'' are as small as
possible. This is usually achieved when lines of dipoles form closed
loops.
The orientation of the spin is determined by its angle $\phi$.
For example, when the azimuth angle $\theta$ can be
described as a vortex with $\phi=\theta-\pi/2$ the dipolar energy is minimized.
When this type of global constraint is satisfied, the domain structure
is usually determined by minimizing the exchange and the anisotropy energy;
we obtain the equation:
\begin{equation}
\nabla ^2 \phi-0.5K \sin 4\phi/{\bar J}=0, \label{dwe}
\end{equation}
Here ${\bar J\approx zJ/4}$ is the
effective exchange.
z is the number of nearest neighbours. It comes from converting the
discrete model to the continuum approximation.
The exactly soluable sine-Gordon equation
$(\partial_x^2-\partial_t^2)\phi-0.5K\sin 4\phi/{\bar J}=0$ is
formally the same as
the above equation (\ref{dwe}) if we transform the $y$
coordinate into the imaginary time $it$.
In this way, we can generate a 90 degree domain wall
``soliton'' solution as $\phi=\tan^{-1} \exp [-\sqrt{2K/{\bar J}}x]$
where the angle $\phi$ changes by 90 degree as the wall is traversed
and x changes sign.
This solution is one dimensional and is well known \cite{Kittel}.

{\bf Many} soliton solutions are known but have never been exploited in
the understanding of domain structures.
A general two soliton solution of the sine-Gordon equation has the
form (\cite{zakh}):
\begin{equation}
\phi=\tan^{-1}\left\{\frac{1-\frac{1-u_1u_2-\sqrt{(1-u_1^2)(1-u_2^2)}}
{1-u_1u_2+\sqrt{(1-u_1^2)(1-u_2^2)}}
\exp\left[-\frac{x'-x_1'-u_1t'}{\sqrt
{1-u_1^2}}-\frac{x'-x_2'-u_2t'}{\sqrt{1-u_2^2}}\right]}
{\exp\left[-\frac{x'-x_1'-u_1t'}{\sqrt
{1-u_1^2}}\right]+\exp\left[-\frac{x'-x_2'-u_2t'}{\sqrt{1-u_2^2}}\right]}
\right\}.
\label{gtss}
\end{equation}
Here $x'=x\sqrt{2K/{\bar J}}, t'=t\sqrt{2K/{\bar J}}$. This solution has 4
arbitrary constants $u_1, u_2, x_1, x_2$. Using the transformation
$u_1=iv, u_2=-iv, t'=i(y'-y_1'), x_1'=x_2'$ and choosing $x_1'=(\ln (1/v)
-i\pi/2)/\gamma + x_0', y_1'= -i\pi/(2 v \gamma)+ y_0'$, we obtain the two
soliton solution in the form:
\begin{equation}
\phi=\tan^{-1} [\sinh(\gamma v\sqrt{2K/{\bar J}}(y-y_0))/(-v
\sinh(\gamma\sqrt{2K/{\bar J}}(x-x_0)))], \label{vort}
\end{equation}
where $v$ is a parameter, $\gamma=1/\sqrt{1+v^2}$.
This describes a closure domain. An example is shown in Fig. 1B
for a triangular lattice of 3600 spins for $K=0.2$ and $J=2$.
A closure domain can be viewed as the space-time trajectory of two
solitons coming together and eventually moving apart.
The parameter $v$ describes the orientation of the domain
wall. To fit into a sample of aspect ratio r, one expects $v=r$, as we
have verified directly by numerical calculation. For a triangular
lattice, the center of the defect, $(x_0, y_0)$
for the lowest energy configuration sits in the middle of the triangle.
This type of domain wall is often
observed in simulations in systems in zero external magnetic field. A typical
finite temperature
simulation result\cite{2l} is also shown in Fig. 1A for the same value of J and
K and $g=1$, obtained from cooling a high
temperature configuration that starts off with all spins aligned in the
$x$ direction.
To study the possible effect of the dipolar interaction
and the accuracy of the analytic formula, we have
numerically minimized the {\bf total} energy of the system
starting from the configuration given by the analytic formula
and using a quasi-Newton algoraithm for a system with 400 spins.
We have explored different values of g less than 1
and find that the mean square difference between
the initial and final azimuthal angles is less than 0.1 radian, out of a
possible range of $\pi$. Thus the accuracy is 3\%;
the analytic formula is indeed a good
approximation. With this analytic formula, it is much easier to
investigate the physical properties of closure domains {\bf quantitatively}.
For example, we have investigated the size dependence of the
energy difference between the closure
domain and that with uniform magnetization along the $x$ direction.
The difference in energy divided by the effective coupling constants (g
for the dipolar energy and $\sqrt{JK}$ for the sum of the exchange and
the anisotropy energy)
is shown in Fig. 3 as a function of the sample size.
For a rectangular sample of a triangular lattice with an aspect ratio
of 0.866 and $x$ dimension $L_1$, the dipolar energy difference
$\Delta E_p$ can be fitted
by the formula $g(109.5-10.54L_1)$ with an error of less than 4\%
whereas the sum of the anisotropy
and exchange energy $\Delta E_w$ can be fitted by the formula
$\sqrt{JK}(28.76+2.64L_1)$ with an error of less than 0.3\%.
The closure domain is lower in energy than the uniformly magnetized
state when the sum of these two
energies become negative. For a film of thickness d, we expect that
approximately $g=g_0d^2$, $J=J_0d$, $K=K_0d$ where the subscript 0
refers to the bare coupling per spin.
Thus the closure domain is lower in energy for
sample sizes $L_1>(109.5+2.876 \sqrt{J_0K_0}/g_0d)
/(10.54-2.64\sqrt{J_0K_0}/g_0d)$. This can only happen if the
denominator is positive; ie $d>d_c=0.25\sqrt{J_0K_0/g_0}$.
As an example, consider bcc Fe
where $g_0\approx 0.254 K$, $K_0\approx 0.038 K$, $J_0\approx 500K$.
Thus the closure domain is lower in energy for thicknesses $d>4.3$ layers.

The solutions (\ref{vort}) is, strictly speaking, applicable
to infinite samples. The consideration of the domain patterns in small
structures require the imposition of finite boundary conditions. The solutions
of Eq. (\ref{dwe}) which satisfy these boundary conditions can be obtained
starting from the ansatz suggested by Lamb \cite{lamb} for the solution of
the sine-Gordon equation. We seek solutions of Eq. (\ref{dwe}) having the
form $\phi(x,y)=\tan^{-1}[f(x')g(y')]$,
where $f$ and $g$ are, in general, Jacobian elliptic functions defined by
\cite{byrd}
$(f')^2=\alpha f^4+\beta f^2-\gamma$ and $(g')^2=-\gamma g^4-(\beta-1) g^2
+\alpha$ with $\alpha, \beta$, and $\gamma$ arbitrary constants,
$x'=\sqrt{2K/{\bar J}}x$, $y'=\sqrt{2K/{\bar J}}y$. As an example, we
consider configurations corresponding to edge domains with the boundary
conditions that the spins point up (down) on the left (right) edge and
horizontally on both the top and the bottom edge.
\begin{equation}
\phi=\tan^{-1}\left[A {\rm tn}(\Omega x', \lambda_f)
\frac{{\rm cn}(v \sqrt{1+k_g^2} y', k_{1g})}
{{\rm dn}(v \sqrt{1+k_g^2} y', k_{1g})}\right], \label{edge}    
\end{equation}
where $k_g^2=[A^2\Omega^2(1-A^2)]/[\Omega^2(1-A^2)^2-1],$
$k_{1g}^2=A^2\Omega^2(1-A^2)/(\Omega^2(1-A^2)-1)$,
$\lambda_f^2=[A^2+\Omega^2(1-A^2)^2]/[\Omega^2(1-A^2)]$ and 
$v^2=[\Omega^2(1-A^2)^2-1]/[1-A^2]$.
The parameters $A$ and $\Omega$ can be determined by requiring that the
component of {\bf S} normal to the surface boundary be zero so that
the dipolar energy is minimized.

Figure 2B 
shows the edge domain pattern obtained by using Eqs. (\ref{edge})
for a triangular lattice 3600 spins for $J=2$ and $K=0.2$.
In Figure 2A we 
show the Monte Carlo result\cite{2l} for a bilayer system for a triangular
lattice of 3600 spins for the same value of J and K and g=1.
Similar domain patterns are also seen in the zero field remanent state
for a system with a single layer.\cite{nuclang}

To study the possible effect of the dipolar interaction
and the accuracy of the analytic formula, we have
numerically minimized the {\bf total} energy of the system 
starting from the configuration given by the analytic formula
and using a quasi-Newton algoraithm for a system with 400 spins. 
When the dipolar interaction is too small, our algorithm recovers the
minimum energy state of uniform magnetization.
As long as the dipolar interaction
is big enough the minimum energy configuration from our algoraithm
is essentially independent of
the strength of the dipolar interaction.
We obtain a state that resembles our analytic results.
The mean square difference between
the initial and final azimuthal angles is less than $10\%$.
The analytic formula is indeed a good
approximation, even though it is not as good as that for the closure
domains. With this analytic formula, 
we have investigated the size dependence of the
energy difference between the edge 
domain and that with uniform magnetization along the x direction.
The results are shown in Fig. 3.
For a rectangular sample of a triangular lattice with an aspect ratio 
of 0.866 and x dimension $L_1$, the dipolar energy difference
$\Delta E_p$ can be fitted 
by the formula $g(52.87-3.97L_1)$ with an error of less than 5\%
whereas the sum of the anisotropy
and exchange energy $\Delta E_w$ can be fitted by the formula 
$\sqrt{JK}(10.46+1.9L_1)$ with an error of less than 1\%.
The edge domain is thus of lower energy when the sum of these two
energies become negative. As expected, when compared with the closure
domains, the dipolar energy gained is less while the cost in the
anisotropy and exchange energy is also smaller. For a film of thickness d,
the edge domain is lower in energy for
sample sizes $L_1>(52.87+10.46 \sqrt{J_0K_0}/g_0d)
/(3.97-1.9\sqrt{J_0K_0}/g_0d)$. This can only happen if the
denominator is positive; ie $d>d_{ec}=2\sqrt{J_0K_0/g_0}$.
For bcc Fe, $d_{ec}=8.2$ layers.

In this paper we have discussed two examples of analytic solutions for
domain patterns. Many possibilities remain to be explored.
For example, consider
\begin{equation}
\phi=\tan^{-1} [\cos(\gamma v\sqrt{2K/{\bar J}}(y-y0))/(v
\sinh(\gamma\sqrt{2K/{\bar J}}(x-x_0))]. \label{tdw}
\end{equation}
where   $\gamma=1/\sqrt{1-v^2}$.
This solution can be considered the analytic continuation of the 
solution (\ref{vort}) with an imaginary $v$.
When $v$ is small, this solution describes two 90 degree domain
walls separated by a distance $2\ln(2/v)\sqrt{{\bar J}/2K}/\gamma)$.
As $v$ is increased from zero, two separated 90 degree domain wall
merge to become a 180 degree domain wall with vortices in between.
This type of solutions is not the lowest energy configuration in zero
magnetic field but occurs as a rate limiting step in spin reversal
processes at a {\bf finite} magnetic field. Our solution provides for
configurations that are local extrema of the exchange and anisotropy
energy. The ordinary 180 degree domain wall in {\bf zero} field, which conisists
of two 90 degree domain walls, is {\bf not} a local extrema of the
exchange and anisotropy energy. It is only stabilized by the magnetoelastic
or dipolar energy.\cite{ber1}

In summary, we have provided examples of how the many soliton solutions
can be used to understand the domain structures in ultra-thin films.
This opens the door to analytic quantatitive understanding of the
micromagnetics in these systems.

This work is supported in part by the Office of Naval Research under
contract N00014-94-1-0213. VNR acknowledges the financial support from the 
Russian Science Foundation through the Grant N96-02-16211 and the 
hospitality of the Bartol Research Institute.

%\bibliography{/home/chui/wp/mag/mag}
%\bibliographystyle{/home/chui/wp/tex/macros/prsty}

\begin{figure}
\caption{Closure domain configuration for a rectangle from a 2 soliton
solution. The analytic results is in B. The finite temperature
Monte Carlo results observed
in ref. 11 is shown in Fig. A.}
\end{figure}
\begin{figure}
\caption{Edge domain configuration for a rectangle from a 2 soliton
solution. The analytic results is in B. The finite temperature
Monte Carlo results observed
in ref. 11 is shown in Fig. A.}
\end{figure}
\begin{figure}
\caption{The energy difference between the domain configuration and that
with uniform magnetization as a function of the linear dimension of the
sample. These energy differences are normalized by
the coupling constants, as is described in the text.}
\end{figure}
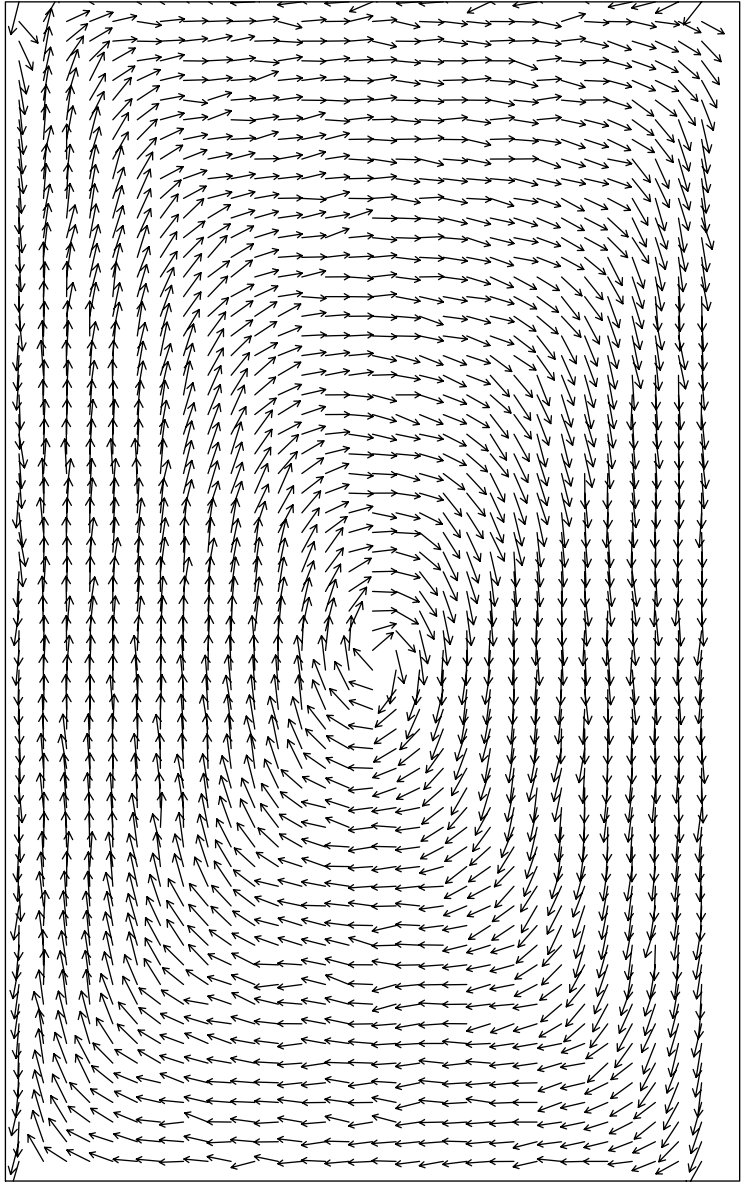
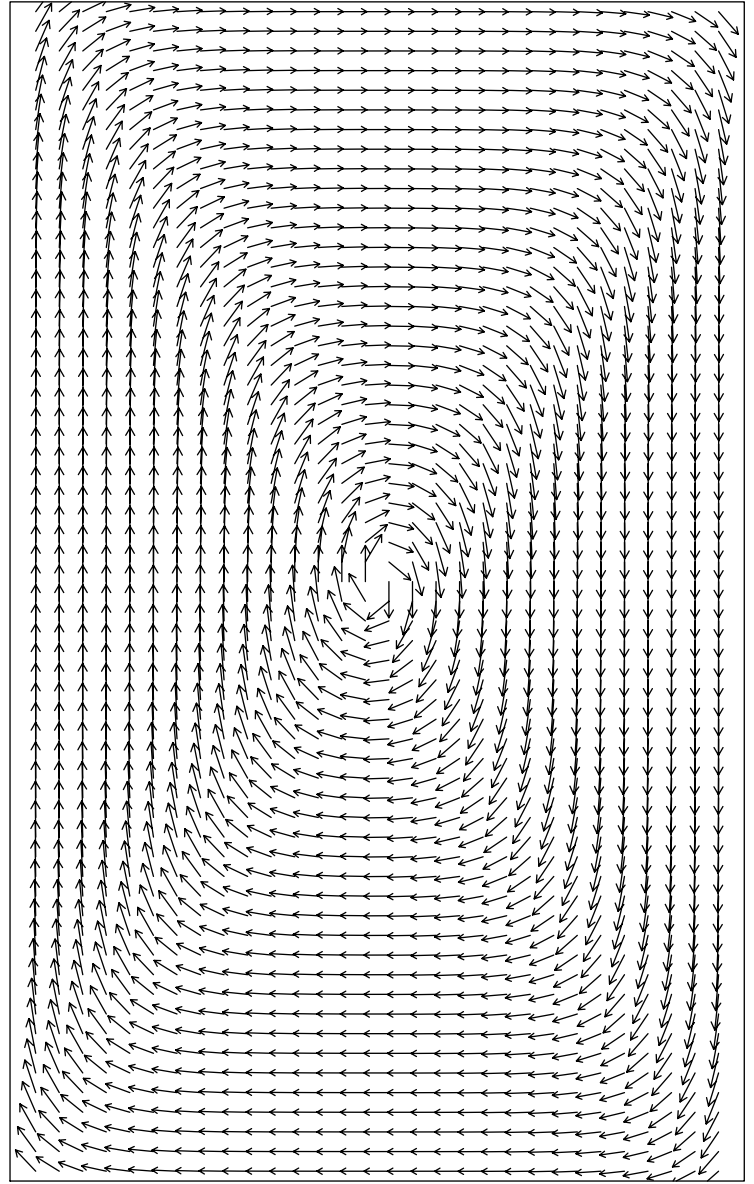
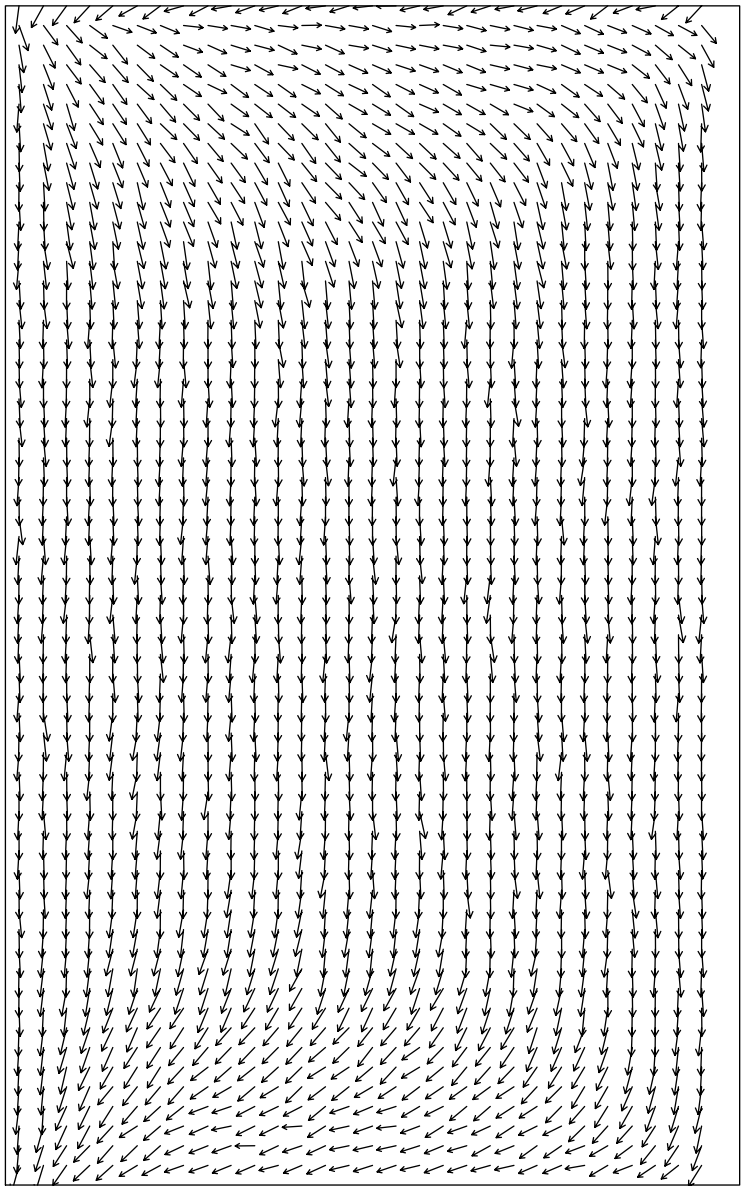
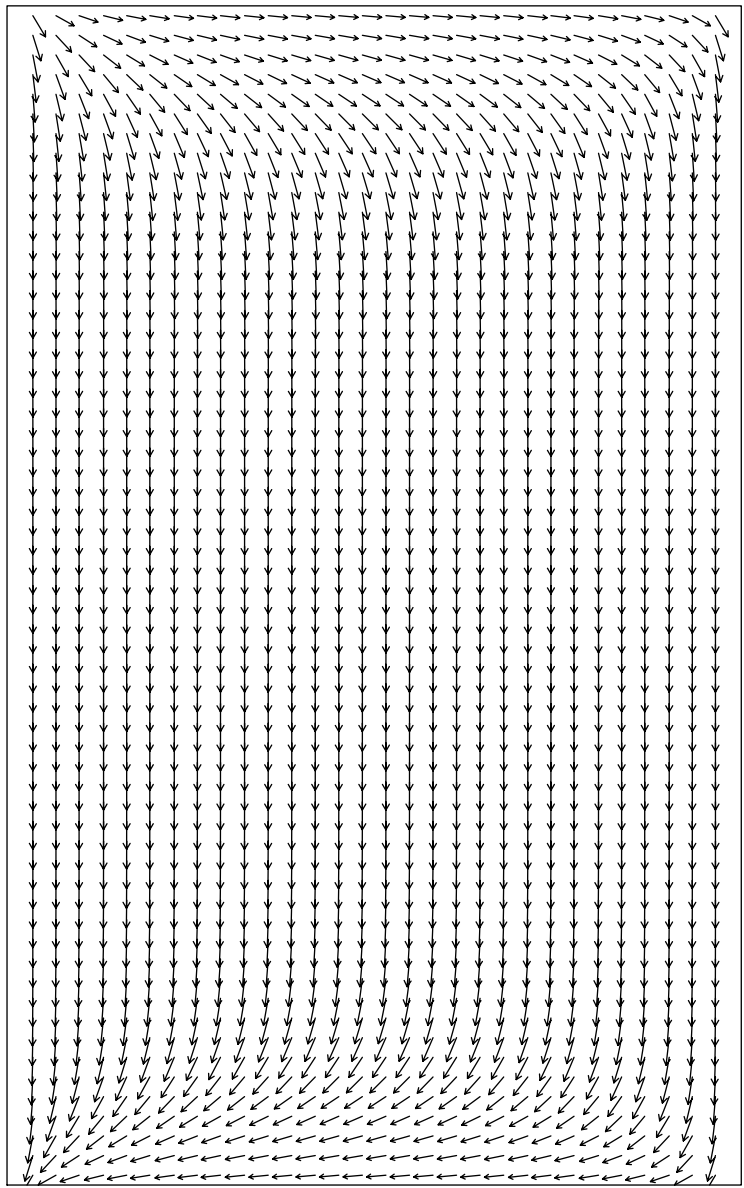
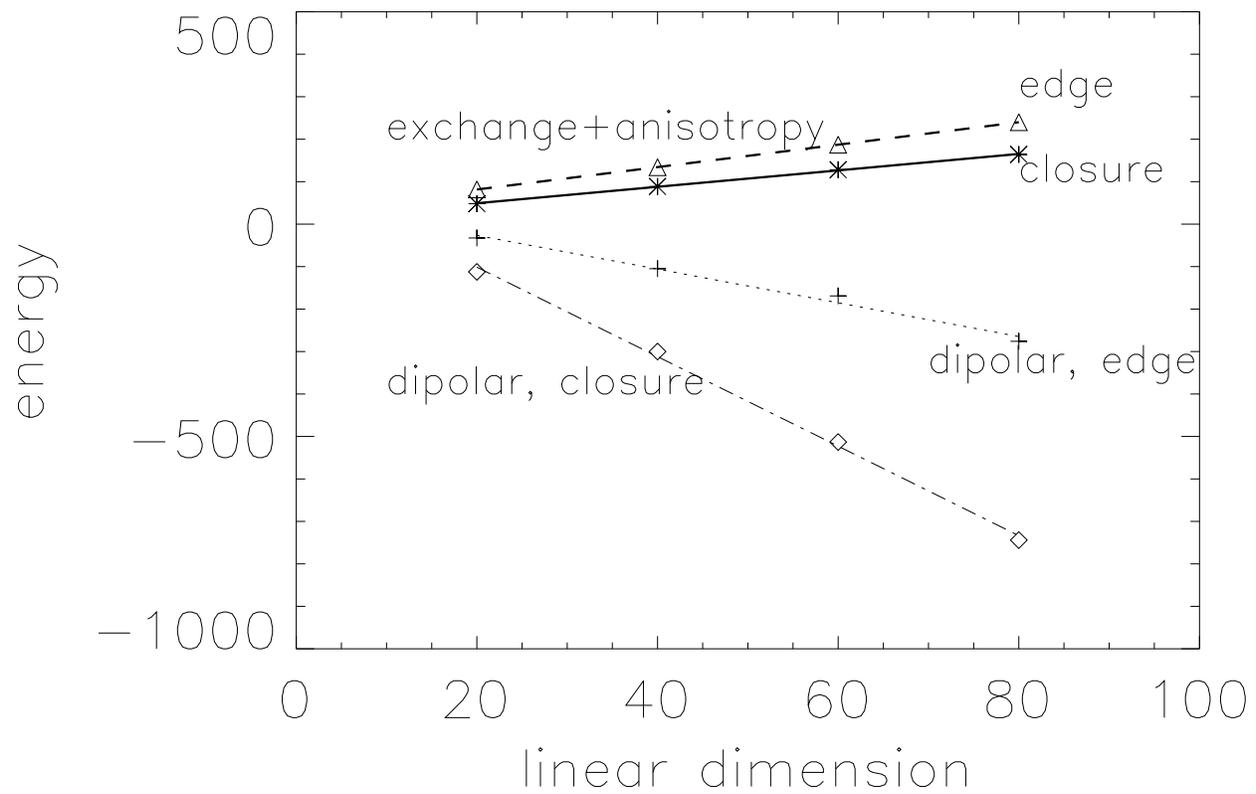
\end{document}